# CB2CF: A Neural Multiview Content-to-Collaborative Filtering Model for Completely Cold Item Recommendations

Oren Barkan
Microsoft, Israel

Noam Koenigstein
Microsoft, Israel
Tel Aviv University, Israel

Eylon Yogev
Microsoft, Israel

Ori Katz
Microsoft, Israel
Technion, Israel

## ABSTRACT

In Recommender Systems research, algorithms are often characterized as either Collaborative Filtering (CF) or Content Based (CB). CF algorithms are trained using a dataset of user preferences while CB algorithms are typically based on item profiles. These approaches harness different data sources and therefore the resulting recommended items are generally very different. This paper presents the CB2CF, a deep neural multiview model that serves as a bridge from items content into their CF representations. CB2CF is a "real-world" algorithm designed for Microsoft Store services that handle around a billion users worldwide. CB2CF is demonstrated on movies and apps recommendations, where it is shown to outperform an alternative CB model on completely cold items.

## KEYWORDS

Cold item recommendations, multiview representation learning, deep learning, recommender systems, item recommendation, collaborative filtering, content based filtering, cold start, natural language understading

## 1. Introduction

Windows Store is powered by Microsoft's recommendations services that cater around 1 billion users worldwide (including MS store and Xbox). The store features many different types of recommendation scenarios, and the "similar items list" (titled "People Also Like") is a principal driver of user engagement and sales. When a new (cold) item is introduced, we wish to be able to find its most similar items based on the CB data alone. Hence, this paper presents the CB2CF algorithm for learning *similar item relations*. CB2CF aims to produce item similarities from content alone where supervision for training the model comes from an external CF model.

In Recommender Systems research, CF models are commonly used for a variety of personalization tasks [9–11]. A common approach in CF is to learn a low-dimensional latent space that captures the user's preference patterns or "taste". For example, Matrix Factorization (MF) models [8, 12, 44] are usually used to map users and items into a dense manifold using a dataset of usage patterns or explicit ratings. An alternative to the CF approach is the Content Based (CB) approach which uses item profiles such as metadata and item descriptions, etc. CF approaches are generally accepted to be more accurate than CB approaches as they capture specific user taste [32].

CB2CF enables the embedding of **completely** cold items (items for which usage patterns do not exist) in the CF space. This capability is crucial in real world recommendation systems in which new items are being added to the catalog on a daily basis, with zero usage information. Typically, these new items are associated with CB profiles that are obtained from a mix of categorical, continuous and unstructured textual data. For example, the CB data of a movie contains tags (e.g. genres), numeric values (release year) and textual description (plot).

CB2CF is trained using a set of items for which both CB and CF data exist. The CB representation of an item consists of multiple CB sources and the CF item vectors are produced by the Bayesian Personalized Ranking (BPR) [12] algorithm. Then, CB2CF learns a multiview mapping from the CB item representation (the input) to its corresponding CF item vector (output).

CB2CF utilizes a convolutional neural network (CNN) on top of word2vec [5] representation to learn a mapping from the textual description of items (movie plots or app descriptions) to their CF item vectors produced by BPR. Beyond the textual descriptions, the model can be enhanced by adding different types of structured metadata as input. This metadata, when used along with the textual information, can be used to improve the overall accuracy of the CB2CF model.

We note that we chose to use the highly popular BPR model for better reproducibility. However, the choice of supervision is configurable and can be replaced by a variety of other CF methods [8]. Similarly, our choice of CNN [7] for processing the CB data can be replaced with any other model such as in [36, 37]. The primary emphasis of this paper is on investigating the connection between items' CB and their CF representations. CB2CF does not allude to compete with existing CF or hybrid models.

The contribution of this work is twofold: First, we introduce the CB2CF model for bridging the gap between items' CB profiles and their CF representations. This is particularly useful for recommending new items for which usage data is not available (cold-start). We show that CB2CF, which is supervised by CF information, produces significantly better results than classical CB models that use the **same** CB data. Secondly, we present a multiview architecture that supports a combination of categorical, continuous and unstructured data as inputs. Finally, we investigate the contribution of each content information source with respect to the CF prediction task and show their ability to reveal interesting patterns that exist in CF datasets. This last trait, although not the main focus of this paper, has the potential to be further harnessed to improve recommender systems expandability.

## 2. Related Work

Deep learning models are being applied in a growing number of machine learning applications. Considerable technological advancements have been achieved in the fields of computer vision [1] and speech recognition [2]. In Natural Language Processing (NLP), deep learning methods have been mostly focused on learning word vector representations [3-6,30]. Specifically, Skip-Gram with Negative Sampling (SGNS) [5], known also as word2vec, has drawn much attention for its versatile uses in several linguistic tasks.

Word2vec maps a sparse 1-of-V encoding (where V is the size of the vocabulary) into a dense low-dimensional latent space which encodes semantic information. The resulting word representations span a manifold in which semantically related words are close to each other. A recent work by Kim [7] has further enhanced this approach by applying a CNN on top of the latent word representations to glean more information from unstructured textual data.

The first part of our model starts from a similar architecture: First, a word2vec model is established in order to map words taken from the item descriptions into a latent semantic manifold. Then, a CNN model is placed in cascade in order to utilize the semantic information for predicting the CF representation of the items. Therefore, the model in this paper serves as a mapping between the content profiles of items and their CF representations.

Many attempts have been taken to leverage multiple views for representation learning [25]-[28]. In the context of Recommender Systems, Wang et al. [20] proposed a hierarchical Bayesian model for learning a joint representation for content information and collaborative filtering ratings. Djuric et al. [21] introduced hierarchical neural language models for joint representation of streaming documents and their content with application to personalized recommendations. Xiao and Quan [22] suggested a hybrid recommendation algorithm based on collaborative filtering and word2vec, where recommendations scores are computed by a weighted combination of CF and CB scores. Other hybrid recommender systems are introduced in [38]-[40].

This paper differs from the aforementioned works by several aspects: First, we focus on the scenario of **completely** cold items, namely items for which CF information is completely unavailable. For such items, a joint representation cannot be inferred, and recommendations are restricted solely to CB data. Hence, CB2CF is trained to infer the CF view of items directly from their CB view. Second, CB2CF introduces a flexible model architecture to support a combination of various input types simultaneously. Third, CB2CF is tasked with learning the CF representation of items only and does not attempt at learning user representations.

Arguably, the most similar previous work is [35], where the authors proposed to learn the song representations using their audio tracks alone. Different from [35], we present a multiview model that supports a variety of different input types. A direct comparison with [35] is redundant: Since the models are based on completely different CB inputs, a comparison would be informative only to the level of which the CB inputs are useful with respect to the task at hand. We do however provide an evaluation for CB2CF that is far more extensive than [35]. We cover two different domains (movies and apps) and an analysis of the contribution of each information source (view).

## 3. The CB2CF Model

Next, we provide a detailed description of the proposed CB2CF model. Recall that our task is to predict the CF representation for each item from its content (textual description / metadata). This work focuses on inferring item relations from implicit feedback only [12]. Given a finite set of items $I = \{k\}_{k=1}^{K}$, users $U = \{u\}_{u=1}^{Q}$ and a user-item matrix $A \in \{0,1\}^{Q \times K}$, we first employ BPR [12] to produce a mapping $M_{CF}: I \to \mathbb{R}^n$ from an item $k$ to its corresponding CF vector $v$. Note that we discard the user vectors obtained by the application of BPR to $A$, as we are interested in the item vectors only.

The CB representation for an item $k$ is obtained by using different mappings for different information sources. For the textual descriptions, we denote a mapping $M_{w2v}: I \to \mathbb{R}^{l \times m}$ from an item to a matrix that consists of $l$ rows, where each row is a $m$-dimensional word vector obtained by word2vec [5]. This matrix corresponds to the first $l$ words in the textual description of the item. If the number of words is less than $l$, we pad the matrix with zero vectors. $M_{w2v}$ is used to generate the input for the CNN-based text component.

The second mapping for textual data maps an item to its bag of words (BOW) representation denoted by $M_{BOW}: I \to [0,1]^b$. This mapping is obtained by first applying k-means clustering on the word2vec representations of the entire vocabulary. We denote the number of clusters by $b$. Then, given the item's description text, soft alignment is applied between each word vector in the text and the $b$ centroids. The result is a histogram vector, which is then normalized into probabilities to form the BOW representations. This approach is inspired by prominent BOW models in computer vision [15] (an alternative is using TFIDF [16], however, in our initial experiments the BOW approach outperformed TFIDF).

For CB information in the form of tags / categories, we define a mapping $M_{tags}: I \to \{0,1\}^T$ from an item to a binary vector, where $T$ is the size of available tags. Each entry in the binary vector indicates whether the tag is associated to the item or not.

The last mapping we apply is used for numerical inputs and denoted by $M_{num}: I \to \mathbb{R}^c$. In this work, the numeric feature is a movie's release year solely. Therefore, in this case $M_{num}$ reduces to $M_{num}: I \to \mathbb{N}$.

In order to harness the different information sources, we utilize a neural multiview regression model consisting of three distinct types of components corresponding to each type of information source: textual, tags and numeric information. In what follows, we describe this architecture in detail.

### 3.1 Text Components

The text components are designed to receive raw text as input and output a fixed size vector. In this work, we implement two different types of text components: dubbed 'CNN' and 'BOW' (marked red in Fig. 1). The CNN approach follows Kim's 'CNN non-static' model from [7]. As explained earlier, using $M_{w2v}$, we map the sequence of words in the textual input to a matrix which serves as the input to a CNN network. An illustration of this approach (taken from [7]) appears in Fig. 1(d). We note that the backpropagation process continues through the CNN onto the initial word2vec representations allowing the word embedding to freely adjust with respect to the CF prediction task at hand. Hence, the initial mapping $M_{w2v}$ is fine-tuned throughout the training process.

Our CNN consists of a single 1D convolutional layer with a filter length of 3 and L2 regularization. This is followed by a global max pooling layer (convolution and pooling are applied over the time axis) and an additional fully connected (FC) layer. In contrast to [7], we did not apply parallel convolutional layers with different filter lengths. We did experiment with multiple filter lengths (2-12), but these attempts failed to materialize into any gains with respect to our objective. A similar observation was also made in [31].

We applied a random dropout of words in front of the CNN. This technique was instrumental in improving the model's generalization capability and avoiding overfitting. The probability for dropping words can be either fixed or proportional to the words' frequency. We found that both methods yield similar results and therefore resulted to using a fixed dropping probability of 0.2.

We considered several additional variants of CNN-based models as in [7]: (1) the 'CNN random' model learns the word representation $M_{w2v}$ from scratch by using random initialization of the word vectors; (2) the 'CNN static' model that keeps the word2vec representation $M_{w2v}$ fixed during the entire training process; (3) the 'CNN multichannel' model, which is a combination of both the 'non-static' and 'static' models. However, the 'CNN non-static' variant outperformed all the rest. In the remainder of this paper, we refer to this variant as our CNN component.

Our second approach for utilizing textual information is based on BOW on top of the word2vec representations. The BOW representation is computed by $M_{BOW}$ and is fed into a neural network with two FC layers and dropout in between. The BOW network architecture is presented in Fig. 1(c).

## 3.2 Tags Components

Beyond the textual information, it might be useful to utilize tags metadata which are associated with each item. The tags network consists of a binary input layer in the dimension of number of tags followed by a single FC hidden layer with L2 regularization on its weights. No further improvement was gained by including additional layers. The input for the tags component is given by $M_{tags}$. The tags component is illustrated in Fig 1(a).

In the movies examples, we used different tags components for different types of metadata: genres, actors, directors and language tags. The hidden layer dimension is determined for each component according to the number of tags and their available combinations. For example, the actors component might be assigned with a higher output dimension than the language component. This is due to the fact that the number of actors is much larger than the number of languages. Moreover, movies usually contain multiple actors, but a single language.

## 3.3 Numeric Components

Numeric components are designed to handle numeric structured data represented as continuous feature vectors. In this work, the only numeric values available were the movies' release year. Therefore, the numeric component was simply chosen to be a network with a single input neuron (Fig. 1(b)). This input is given by$M_{num}$.

## 3.4 The Combiner Component

The combiner component aims at combining multiple outputs from different components in order to predict into the CF space. The combiner component (illustrated in Fig. 1(e)) consists of multiple input layers that are fully connected to a hidden layer with L2 regularization. Hence, the combiner **concatenates** all outputs from previous layers to form a single layer that is followed by a final FC output layer in the same dimension of the CF space.

## 3.5 The Full CB2CF Model

The CB2CF model is illustrated in Fig. 2. In accordance with Fig. 1, tags, text and numeric components are colored in blue, red and green respectively. The combiner component is colored in yellow. Fig. 2 exemplifies the application of the presented model for the movie similarity task, specifically for the 'Pulp Fiction' movie. Genres, actors, directors and languages are modeled as tags components, the movie plot summary is modeled as a text component. In this implementation, the text component can be either CNN or BOW network. The movie's release year is modeled as a numeric component. All of the components' outputs are then fed into the combiner component that outputs the predicted CF vector. The loss function we use to train the model is the Mean Square Error (MSE), which is a common choice for regression tasks. ReLU [1] activations were found to outperform other activation functions such as in [43] and are used in between model layers. In inference phase, the affinity between items is computed via the **cosine similarity** of their respective vectors.

It is important to emphasize that the proposed CB2CF model is extremely flexible in the sense that each component can be easily disconnected from the combiner and the extension for additional information sources is straightforward. For example, we can add the countries where the movie was filmed and the movie duration as additional tags and numeric components, respectively. In addition, other advanced NLP models such as in [36], [37] can be used as text components, however we leave these models for future investigation. Finally, the exact hyperparameters configuration is detailed in Section 4.3.

# 4. Experimental Setup and Results

The quantitative results in this paper are obtained by 10-fold cross validation. We supplement these quantitative results with qualitative results to gain a better "feel" of the model. Recall that our goal is to predict for each item its CF vector from its CB profile. Hence, the CF representation is considered as the *ground truth* when training the CB2CF model. Next, we describe in detail the datasets, evaluated systems, parameter configurations, evaluation measures and present results.

## 4.1 Datasets and Data Preparation

In this section, we present two experiments: The first experiment provides an ablation study of the CB2CF model. The second experiment provides a comparison between CB2CF and other existing methods on completely cold item recommendations.

### *4.1.1 Word2vec dataset*

We used a subset of the dataset from [14] in order to establish a word2vec model. Specifically, we kept only the top 50K most frequent words. We also mapped all numbers to the digit 9 and removed punctuation characters. Then, we randomly sampled 9.2M sentences that form a total text length of 217M words for training the word2vec model [5]. These word embeddings are used in the text components.

### *4.1.2 Movies dataset*

The movies dataset is publicly available and contains both CF and CB data for movies. The CF data is based on the MovieLens dataset [18] containing 22,884,377 ratings collected during 1995-2016 from 247,753 users that watched 34,208 movies. The movies are rated using a 5-star scale with half-star increments (0.5 - 5.0). From each user's rating list, we consider all the movies with ratings above 3.5 as a set of co-occurring movies. This results in 173,266 users (sets) that contain 11,108 unique items (movies) as the effective training data for learning the BPR model [12].

For each movie, we collected metadata from IMDB [33]. Three types of information sources are collected: movie plot (given as raw text), genres / actors / directors / languages (given as tags) and release year (given as a natural number). In the metadata tags, we filtered tags that with less than 5 occurrences resulting in a remainder of 23 genres, 1526 actors, 470 directors and 72 languages.

We created movie CB profiles as follows: First, we represented each movie's plot summary by taking the first 500 words with word2vec mapping. We used zero padding for the plot descriptions shorter than 500 words. Then, the metadata fields from above were added to the movie profiles. Note that some of the movies had missing information. In this case, we set the plot or the missing tags to a special word / tag 'n/a'. Missing values for the release year are set to the mean year of all movies (1993).

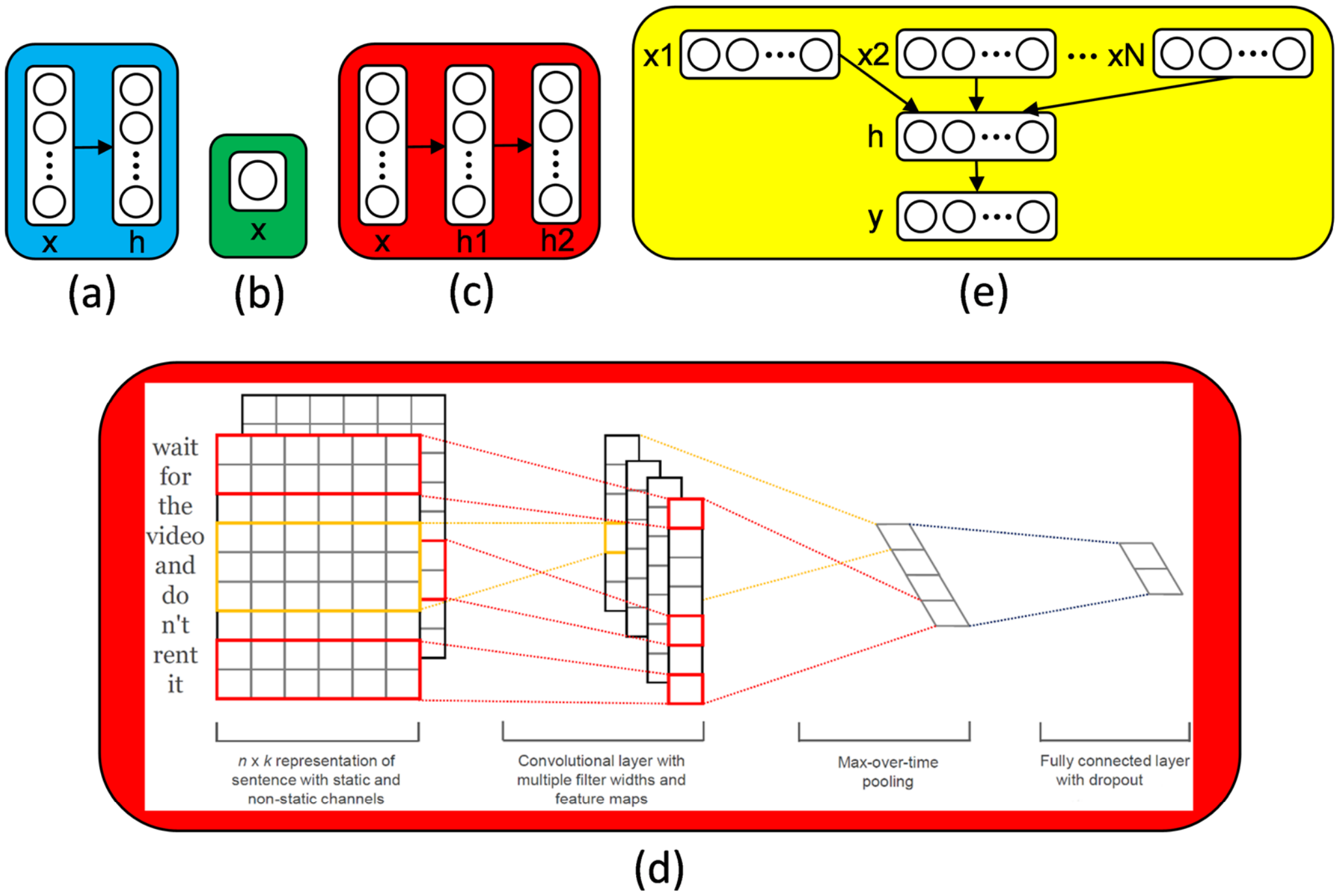

Figure 1. The CB2CF components: Input, hidden and output layers are marked 'x', 'h' and 'y', respectively. Note that each layer may contain different number of neurons. Black arrows symbol FC connections. **(a)** Tags components (Section 3.2). **(b)** The numeric component (Section 3.3). **(c)** The BOW component (Section 3.1) **(d)** The CNN component (Section 3.1) **(e)** The combiner component (Section 3.4) receives the outputs from several different components and fully connects them to a hidden layer 'h' that is followed by the output layer 'y'. The dimension of the output layer is the same as the dimension of the CF space $n$.

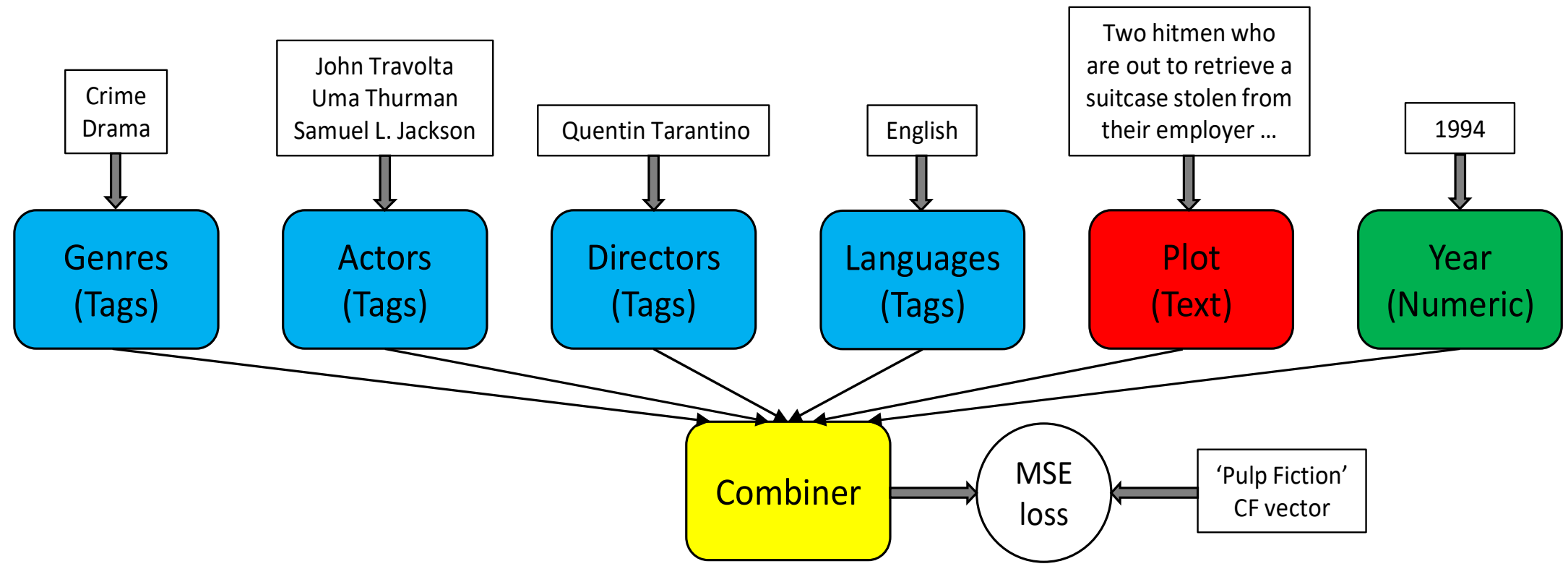

Figure 2. CB2CF model illustration for the movie similarity task. The figure shows an example for the movie 'Pulp Fiction'. Genres, actors, directors and languages are modeled as tags components. Movie plot summary is modeled as text component (can be either CNN or BOW network). The release year is modeled as a numeric component. The combiner receives the outputs from all components and outputs a vector that is compared against the original 'Pulp Fiction' CF vector (produced by BPR) using the MSE loss function.

#### *4.1.3 Windows apps dataset*

The second dataset is a propriety dataset containing CF and CB data for apps from the Microsoft Windows Store. We generated CF profiles for the items using a dataset of users activity containing 5M user sessions. Each user session contains a list of items that were clicked by the same user in the same activity session. This dataset consists of 33K unique items (apps) which were used to procure the BPR model of representative CF vectors.

For each app, we created textual profiles based on the app description in the same manner as we did with the movies data (first 500 words that have word2vec representation are saved for each app as its textual description). In this case, **no** further metadata was used beyond the textual descriptions.

#### *4.1.4 Data preparation*

Since we focus on item recommendations, for each user, we consider each pair of consecutive items $(i, j)$ as a positive example,

where item $i$ is the query and item $j$ is the correct item to be retrieved. We randomly split the users set to training and test users. The training users are used to produce CF item vectors using BPR and the test users are used to form a test set $P$ of consecutive item pairs. In addition, we randomly split the items into two separated sets: the first contains training items and the second contains test items. We denote the test items set by $T$. This procedure was repeated in a 10 fold cross validation manner.

## 4.2 Evaluated Models and Configurations

We consider two BPR models: the first model is 'BPR1' that uses the training users set as is. BPR1 does include items from $T$ in its training procedure and hence is an oracle model that can be treated as an *upper bound* for the CB2CF performance. All of the reported CF results are based on the CF item vectors produced by **BPR1**. However, in reality, a CF model is not exposed to new test cold items during the training phase and cannot infer representations for items without usage data. Yet, using BPR1 as a benchmark does make sense since it is used to quantify the margin between the CB2CF performance and the 'optimal performance'. BPR1 is used for training the different CB2CF in Experiment 1 (Section 4.5).

The second model is 'BPR2' that is trained on a modified training set where for each user all items that belong to $T$ are **omitted**. This ensures that the CF item vectors, produced by BPR, are **not** affected from relations with test items. BPR2 contains CF item vectors for items from the training set **only** and is used for training the CB2CF model in Experiment 2 (Section 4.6). Specifically, we further split the CB2CF training set to training and validation for hyperparameter selection.

The third model is the CB model that maps the items to their BOW vectors based on all information sources. The similarity between the BOW vectors is computed via the cosine similarity.

In order to quantify the relative contribution of each data source in our model, we trained different CB2CF configurations, each time connecting a **single** component to the combiner and **disconnecting** all other components. For tags components we trained separate models for genres, actors, director and language. When presenting results, we intuitively dubbed each of these models according to their information sources i.e., 'Genres', 'Actors', 'Director' and 'Language' respectively. For the text components we trained separate models for CNN and BOW as explained above and dubbed them 'CNN' and 'BOW'. For the numeric component we trained a separate model for the release year and dub it 'Year'. In order to quantify the relative contribution of each combination, we further trained models for the following combinations of components: 'Tags' – a combination of 'Genres', 'Actors', 'Director' and 'Language'. 'Tags+Year' – a combination of 'Tags' and 'Year'. 'Tags+CNN' – a combination of 'CNN' and 'Tags'. 'CNN+Year' – a combination of 'CNN' and' Year'. 'CNN+Tags+Year' – a combination of 'CNN', 'Tags' and 'Year', which is the CB2CF model (Section 3.5). We do not include the 'BOW' component in the combinations since we found its contribution to be marginal once 'CNN' is included.

## 4.3 Hyperparameters Configuration

For all of the evaluated models, hyperparameters were determined according to a separate validation set. The BPR models (for both movies and apps) were trained for 100 epochs with a target dimension $n = 40$. The word2vec model was trained for 100 epochs with a target dimension $m = 100$, window size 4, subsampling parameter 1e-5 and negative to positive ratio 15. For the 'Genres', 'Actors', 'Director' and 'Language' components, we used hidden layers with dimensions 100, 100, 40 and 20, respectively.

The 'CNN' components (for both movies and apps) uses 300 filters with a length 3 (each filter's dimensions are in size of $3 \times 100$). The input shape for the 'CNN' was set to a matrix in size of $500 \times 100$. This matrix contains the first 500 words from the movie plot / app description, where each word vector is in dimension 100. For the 'BOW' component, we used hidden layers of dimension 256. The number of centroids in k-means was set to $b = 256$. For the combiner component, we used a hidden layer of dimension 256.

Each system was trained to minimize the MSE loss function. We used the Adam optimizer [17] with a mini-batch size of 32 and applied an early stopping procedure [19]. When applied, the L2 regularization and dropout probability values were set to 3e-5 and 0.2, respectively.

## 4.4 Evaluation Measures

The first evaluation measure we use is the Mean Squared Error (MSE) as measured by the mean squared difference of the (CB2CF) predicted vectors from their (BPR) CF item vectors. Formally, MSE is measured as follows: $MSE = \frac{1}{|\mathcal{T}|}\sum_{i \in \mathcal{T}} \|v_i - \hat{v}_i\|^2$, where $\mathcal{T}$ is the set of all test set items, $v_i$ is the original CF vector, and $\hat{v}_i$ is the predicted vector. Minimizing the MSE is the objective of all the (CB2CF) systems in this paper. It quantifies the ability of the different systems to reconstruct the original CF vectors produced by BPR. However, MSE does not have any direct business interpretation with regard to the ultimate CF task. Hence, our next evaluation measures are borrowed from the field of CF research and directly quantify the quality of the predicted vectors with regard to the CF task.

The second evaluation measure is the Top-K Mean Accuracy. The Top-K Accuracy function outputs 1 if for a given test item (query), the correct item is ranked among the top k items predicted by the model, otherwise 0. Then, the Top-K Mean Accuracy is obtained by taking the mean across all Top-K accuracies that are computed for all queries.

The third evaluation measure is the Mean Percentile Rank (MPR) [34]: Given a query item the percentile rank (PR) is the rank that assigned by the model to the correct item (to be retrieved) divided by the size of the item catalog. Formally, we denote by $r_i$ the ranked position of the correct item to be retrieved w.r.t. to the query item, when measured against the other items based on similarity to the predicted vector. For a dataset of $M$ items, the best possible rank is $r_i = 0$ and the worst is $r_i = M - 1$. The $MPR$ measure is computed according to $MPR = \frac{1}{|\mathcal{T}|}\sum_{i \in \mathcal{T}} \frac{r_i}{M-1}$. Note that $0 \leq MPR \leq 1$, where $MPR = 0$ is the optimal value (**the lower the better**) and $MPR = 0.5$ can be achieved by random predictions.

Our final evaluation measure was chosen to quantify the ability of the predicted vectors to maintain the original item similarities. Specifically, we care more about the ability to find the most relevant item to each test item. Hence, we chose to use the Normalized Discounted Cumulative Gain [34] for the top $K$ most similar items or $NDCG(K)$. Note that $0 \leq NDCG(K) \leq 1$ and $NDCG(K) = 1$ indicates a "perfect" prediction for the top $K$ most similar items.

## 4.5 Experiment 1 – CF Space Reconstruction (Ablation Study)

This experiment is designed to quantify the relative contribution of each CB source (and their combinations) to **reconstruction** of the CF space, using the CB2CF approach. In this experiment, we used the CF item vectors that are produced by BPR1. Table 1 depicts the MSE and MPR (x100) values and Fig. 3 depicts NDCG(K) for Movies dataset using the systems described in Section 4.2 (Note that MPR and NDCG measures are based on the similarity scores that are computed between the CF item vectors and their CB2CF based reconstruction). One can see that the evaluation measures are highly correlated. In what follows, we identify common trends across all evaluation measures and provide interpretations of these trends.

First, let us consider the 'BOW' vs. the 'CNN' systems. Both systems are based purely on the movie textual descriptions. Our results show that the 'CNN' approach achieves better reconstruction than the 'BOW' approach. This showcases the ability of the 'CNN' model to benefit from the semantic information encoded in the word2vec representations as well as the ability of the 'CNN' filters to pick up the semantical context encoded by word propagation in the text.

Next, we turn to consider the tags data. As explained in Section 3.2, there are four types of tags based systems: 'Genres', 'Actors', 'Directors', and 'Language' and we consider each system separately. Table 1 and Fig. 3 show that these systems were outperformed by 'CNN' across all measures. This indicates the ability of the 'CNN' system to utilize the textual information beyond these very informative data sources.

The 'Year' system, based on movies release year, outperforms each of the previous systems including the 'CNN'. Clearly, a movie's release year alone cannot make for a good recommender system. Nevertheless, it captures a key pattern in the MovieLens dataset, which is characterized by many users who watch movies with adjacent release dates. Typically, movies are heavily promoted during their release period and much of the viewing patterns recorded in the MovieLens dataset occur during that period. Many users watch multiple movies with close release dates. Hence, a movie's release date explains a very dominant pattern in the dataset. Finally, we turn to consider systems in which different information sources are combined. We notice that each combined model generates a considerable performance boost over its respective systems. Ultimately, the 'CNN+Tags+Year' system (the full CB2CF model) outperforms all the rest by combining all these information sources together.

Table 2 presents MPR and NDCG values obtained by the 'CNN' system (CB2CF) on the apps dataset (the apps datasets contains textual descriptions only). We see that the MPR value obtained by the 'CNN' model is much better (lower) than the best result obtained by the CB2CF model for the movies data (2.35 vs 5.4) that leverages metadata as well. We believe this is due to fact the apps dataset contains more items and training examples than the movies dataset (30K vs 11K) that enables a better fine-tuning of the word vectors with respect to the prediction task.

## 4.6 Experiment 2 – Completely Cold Item Recommendations

In experiment 1, we showed that by using all CB information sources, CB2CF produces the best CF space reconstruction. In experiment 2, we investigate the CB2CF item recommendation accuracy, when compared to a plain CB model and the BPR1. The evaluation is done for **completely** cold items – items for which usage is not available and do **not** participate in the training phase of the models. Note that in this experiment, CB2CF is trained using the CF item vectors that are produced by **BPR2** – a BPR model that was not exposed to the items from the test set $T$ (Section 4.2).

In order to examine the generalization performance of CB2CF for cold items, we conduct evaluations on two different test sets of items pairs: The first set is the 'Train Test' set that consists of all pairs $(i, j)$ in $P$, where $i \in T \vee j \in T$. The second set is the 'Test Only' set, where $i \in T \wedge j \in T$ (both items belong to the test items set $T$). Note that the performance of the BPR1 and CB competitors models should be approximately the same across the abovementioned sets, since BPR1 is exposed to all items in the catalog during training and the CB model is based on BOW representation that utilizes CB data only that is independent of usage relations.

The movies and apps datasets contain 10K and 20K test users and 1100 and 2000 test items, respectively. Using these sets we formed the 'Train Test' and 'Test Only' sets that contain 72K and 93K, 8K and 13K test movies and apps pairs, respectively.

Tables 3 and 4 present MPR values (X100) obtained for each combination of test set and model for movies and apps, respectively. As expected, the best results for all sets are obtained by BPR1 (regarded as the optimum oracle). CB2CF significantly outperforms CB on all sets. We further observe that the difference between MPR values that were obtained by CB2CF for different sets is marginal, which means CB2CF generalizes well. As expected, the MPR scores produced by both BPR1 and the CB model exhibit negligible amount of variation across the different sets.

Tables 5 and 6 present the Top-20 mean accuracy values obtained by the evaluated models on the movies and apps datasets, respectively. We see that the same trends from Tabs. 3 and 4 also exist in Tabs. 5 and 6. This is another evidence for the ability of CB2CF to produce better recommendations than the CB model for completely cold items.

## 4.7 Qualitative Results

Table 7 presents movie recommendations based on nearest neighbor search (with cosine similarity) in the BPR1 and the CB2CF spaces with respect to test queries. All queries contain test set items. The second column presents recommendations produced using the CF item vectors of BPR1. The third column presents recommendations produced by CB2CF ('CNN+Tags+Year') that utilize all information sources. The last column presents recommendations produced by the 'CNN' system that leverages textual description of movies (plots) solely.

Three well-known movies from the test set are considered: 'Shrek' (2001), 'The Hangover' (2011) and 'Gladiator' (2000). We notice a tendency of BPR1 recommendations to prefer popular movies. CB2CF tends to pick recommendations from adjacent years having the same genre / actors and with similar plots. The 'CNN' model produces recommendations that are not restricted to a specific year. Therefore, it contains the recommendations 'Shrek the Third' and 'Shrek Forever After' released in later years (2007 and 2010). The 'CNN' based CB2CF exhibits a similar behavior when recommending the third movie in 'The Hangover' series. This showcases the ability of the 'CNN' component alone to provide competitive recommendations.

Table 8 presents app recommendations produced for the Windows Apps dataset, according to similar setting as in Table 5. The second column presents recommendations that were produced using the BPR1 item vectors. The third column presents recommendations produced by CB2CF (using 'CNN' as the Windows Apps dataset contains text descriptions **only**). CB2CF manages to provide accurate recommendations with respect to the given seed based on the textual description of apps solely.

Tables 7 and 8 are empirical evidences for the ability of the CB2CF approach to produce recommendations that are on par with the **oracle** BPR1 model. Together with the extensive quantitative analysis from Sections 4.5 and 4.6, we conclude that CB2CF provides a good CB based out-of-sample extension for the CF

Table 1: MSE/MPR (x100) values obtained by different model configurations on the Movies dataset

| System | MSE / MPR |
|---|---|
| Language | 23.1 / 40.8 |
| Director | 22.2 / 34.3 |
| Actors | 21.6 / 25.5 |
| Genres | 21.3 / 21.4 |
| BOW | 21.2 / 19.2 |
| CNN | 20.3 / 17.2 |
| Year | 19.8 / 15.4 |
| Tags | 19.2 / 12.4 |
| CNN + Tags | 18.6 / 11.2 |
| CNN + Year | 17.4 / 7.6 |
| Tags + Year | 17.1 / 6.7 |
| CNN + Tags + Year (CB2CF) | **16.5 / 5.4** |

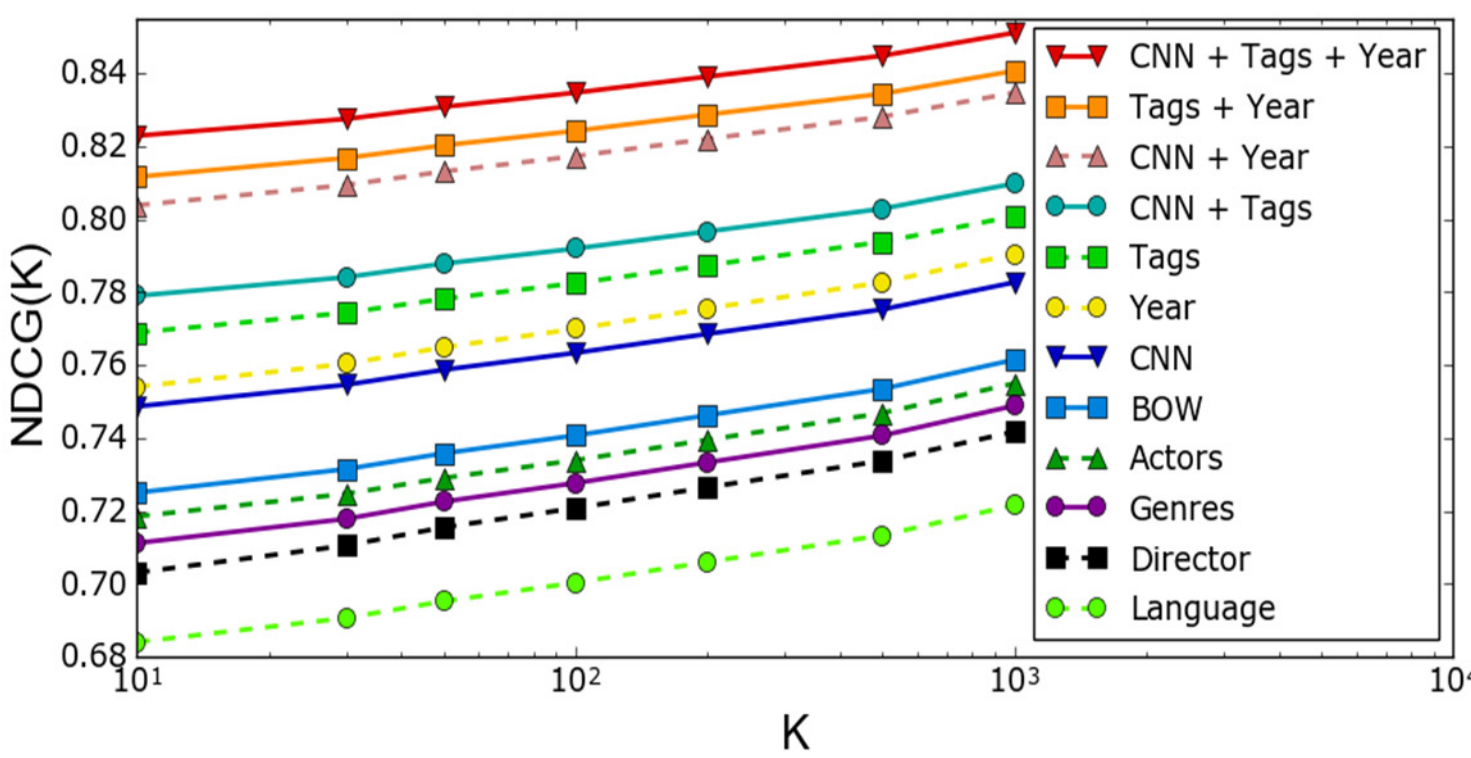

Figure 3. Average NDCG scores obtained by different systems for various K values (10, 30, 50, 100, 200, 500 1000) on 10 fold cross validation. K-axis is in log scale.

Table 2: MPR (x100) and NDCG@10 values obtained on the Apps dataset

| System | MPR / NDCG@10 |
|---|---|
| CB2CF | 2.35 / 0.86 |

Table 3: MPR (x100) values obtained by different models on the Movies test sets

| Model / Set | Train Test | Test Only |
|---|---|---|
| *BPR1 (oracle)* | *15.04* | *15.13* |
| CB2CF | **17.93** | **19.51** |
| CB | 27.18 | 27.23 |

Table 4: MPR (x100) values obtained by different models on the Apps test sets

| Model / Set | Train Test | Test Only |
|---|---|---|
| *BPR1 (oracle)* | *12.69* | *12.56* |
| CB2CF | **14.31** | **15.91** |
| CB | 25.02 | 25.11 |

Table 5: Top-20 mean accuracy values obtained by different models on the Movies test sets

| Model / Set | Train Test | Test Only |
|---|---|---|
| *BPR1 (oracle)* | *0.556* | *0.551* |
| CB2CF | **0.482** | **0.475** |
| CB | 0.372 | 0.369 |

Table 6: Top-20 mean accuracy values obtained by different models on the Apps test sets

| Model / Set | Train Test | Test Only |
|---|---|---|
| *BPR1 (oracle)* | *0.733* | *0.731* |
| CB2CF | **0.616** | **0.598** |
| CB | 0.436 | 0.431 |

Table 7: Top-4 Recommendations produced by different models for test movies

| Query | BPR1 (oracle) | CB2CF using CNN + Tags + Year | CB2CF using CNN only |
|---|---|---|---|
| Shrek (2001) | Monsters Inc., Shrek 2, Finding Nemo, Ice Age | Shrek 2, Stuart little 2, Monsters Inc., Toy Story 2 | Shrek The Third, Shrek Forever After, Shrek 2, Finding Nemo |
| The Hangover (2009) | Superbad, Role Models, I Love You Man, Knocked Up | The Hangover 2, Grown Ups, Role Models, Due Date | 21 jump street, The Hangover 3, The Hangover 2, Grown Ups |
| Gladiator (2000) | The Patriot, The Last Samurai, Saving Private Ryan, Enemy at the Gate | The $13^{th}$ warrior, Story of Joan of arc, The Musketeer, The Last Castle | The $13^{th}$ warrior, King Arthur, 300, Troy |

Table 8: Top-4 Recommendations produced by different models for test apps

| Query | BPR1 (oracle) | CB2CF using CNN only |
|---|---|---|
| Bitcoin Info (Finance) | BitFlow, Bitcoin Blockchain, Bitcoin Values, DogeMuch | Bitcoin Trader, Coin Miner, Bitcoin Markets, Bitcoin Chart+ |
| Pre-League (Soccer) | One Soccer, Soccer Info, La Liga Teams, La Liga | FIFA World Cup'14, The-Football-App, One Soccer, Premier League Hub |
| Cosmetics Magazine (Lifestyle) | Fashion Trends, That Girl, JUSTPROUD Fashion News, Hair and Makeup Artistry | Makeup Tricks, Hair & Beauty, Beauty Tutorials, Natural MakeUp |
| Travel Advice (Travel) | World Destinations, Places to Visit, Local Movies, Animal-Planet | Travel Advisories, Travel Expert, 100 Must See Places, Best Travel Destinations |
| Weight Loser (Fitness) | Ideal Weight, Calculate Your Calories Burned!, 8 for Hourglass, Crunch challenge | Diet Chart for Weight Loss, Calculate Your Calories Burned!, Tips to Lose Weight Fast, Calories Calculator |

Table 9: Semantic relations between movie actors, learned by CB2CF

| | |
|---|---|
| "Dwayne Johnson" - "Sylvester Stallone" = $\Delta_1$ | "Meryl Streep" + $\Delta_1$ = "Cate Blanchett" |
| "Jim Carrey" - "Brad Pitt" = $\Delta_2$ | "Angelina Jolie" + $\Delta_2$ = "Jennifer Aniston" |
| "Richard Gere" - "Hugh Grant" = $\Delta_3$ | "Jason Statham" + $\Delta_3$ = "Vin Diesel" |

manifold and hence can be used as a reliable estimate for the CF vectors of completely cold items.

In the word2vec paper by Mikolov et al. [5], the authors illustrated the ability of their model to automatically organize word representations that capture semantic relations. For example, they showed that the relationship between a country and its capital is captured by the difference between their respective vector representations (Fig. 2 in [5]). Inspired by this, we demonstrate the ability of CB2CF to encode relationships between actors. Actor representations are produced by setting the corresponding entry of the actor in the 'Actors' component to 1 and all other entries to 0.

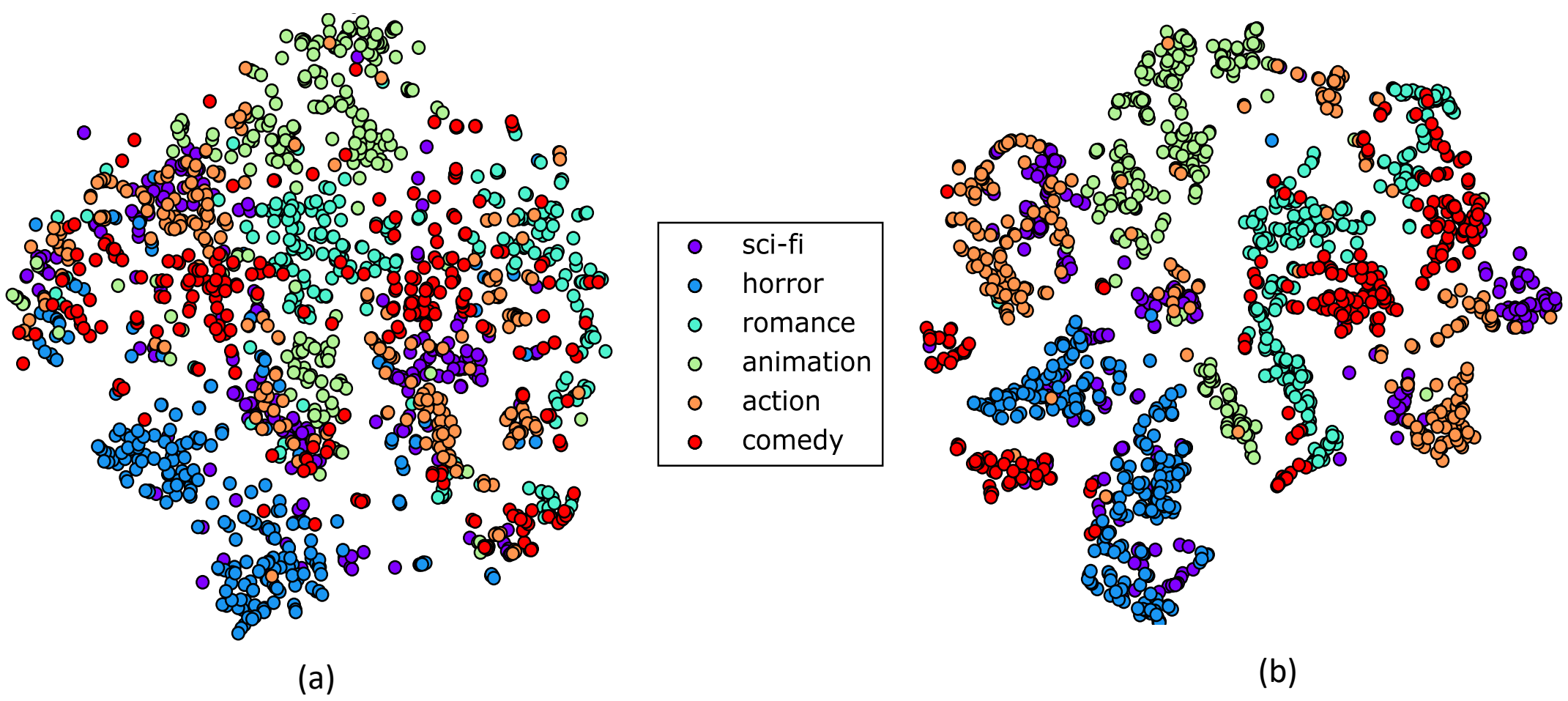

Figure 4: TSNE visualization of the CF item representation produced by BPR1 (a) and the representation produced by CB2CF (b) for 800 test movies. Movies are colored according to their genres.

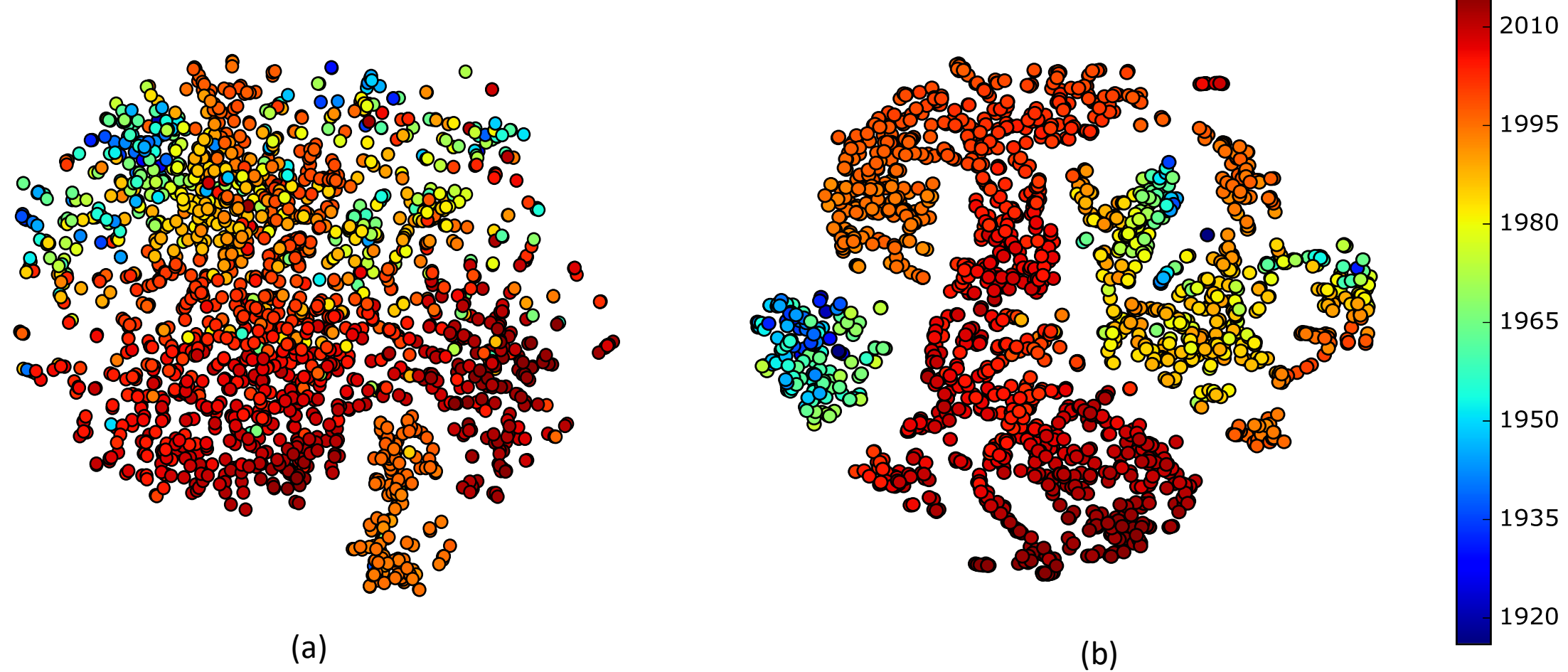

Figure 5: TSNE visualization of the CF item representation produced by BPR1 (a) and the representation produced by CB2CF (b) for random set of 1000 movies. Movies are colored according to their release year.

Table 9 presents the learned relationships. The left column presents a relationship between actors captured by the distance vector between their representations. In the right column, this relationship is applied to a new actor by adding the distance vector to a new origin. The closest artist is then retrieved and presented.

The first example demonstrates a relationship based on a generational gap between actors that play in movies of the same genre. The second example demonstrates a transition from versatile actors to comedy-oriented actors in both genders. Finally, the third example demonstrates a transition from American to British actors across different genres.

Figure 4 depicts a t-SNE [13] embedding of CF item vectors produced by BPR1 (a) and the predicted vectors by the CB2CF model (b) for a random pick of 800 movies from the top six genres. For movies with multiple genres, we simply depict their first tag. The figure indicates that genre clustering exists both in the CF space and even more so in the CB2CF space.

Figure 5 depicts a random pick of 1000 movies from the test set and the t-SNE embedding of their CF vectors produced by BPR1 (a) and the vectors predicted by CB2CF (b). This figure investigates the importance of a movie release dates in finding similar items. In accordance with our evaluations, Fig. 5 indicates that a release date is an important factor in CF similarities as well as in the resulting CB2CF predictions.

## 5. Conclusion

In this paper, we introduced the CB2CF model that enables cold items recommendations via bridging the gap between CB and CF representations. CB2CF learns a neural multiview mapping from the CB space to the CF space and is capable of leveraging various types of CB inputs such as unstructured text, tags and numeric data. We showed that CB2CF outperforms a CB model that is based on BOW (in the word2vec space) in completely cold start scenarios and produces item similarities that are on par with the ones produced by an 'oracle' CF model that is based on BPR.